\documentclass[11pt]{article}
\usepackage{amsmath}
\usepackage{amssymb}
\usepackage{graphicx}
\usepackage{hyperref}

\renewcommand{\baselinestretch}{1.2}
  \renewcommand{\arraystretch}{1.0}
\begin{document}

 \title{On Shor's Factoring Algorithm with More Registers and the Problem to Certify Quantum Computers}

 \author{Zhengjun Cao$^{1,*}$, \qquad Zhenfu Cao$^{2}$}

  \footnotetext{\noindent $^1$Department of Mathematics, Shanghai University, Shanghai, 
  China. \, \textsf{caozhj@shu.edu.cn} \\ 
     $^2${Department of Computer Science and Engineering,  Shanghai Jiao Tong University,  
  China.    \\ $^*$ The extended abstract of this paper has appeared in Proceeding of 2nd International Symposium on Information Science and Engineering, 2009, pp.164-168.}}

\date{}
\maketitle

 \begin{quotation}
 \textbf{Abstract.}
  Shor's  factoring  algorithm  uses two quantum registers. By introducing  more
 registers we show that  the  measured  numbers in these registers which are of the same
pre-measurement state, should be equal if the original Shor's complexity argument is sound.   This
  contradicts the argument that  the second register has $r$ possible
measured values.
 There is an anonymous comment which argues  that the states in  these registers are entangled. If so,
 the entanglement involving many quantum registers can not be interpreted by the mechanism of EPR pairs and the like.
In view of this peculiar entanglement has not yet been mentioned and investigated, we think the claim that the Shor's algorithm runs in polynomial time needs more physical verifications. We also discuss the problem to certify quantum computers.

 \end{quotation}

\section{Introduction}

It is well-known that factoring an integer $n$ can be reduced to finding
the order of an integer $x$ with respect to the module $n$ (G. Miller \cite{M76}),  which is usually denoted by  $\mbox{ord}_n(x).$
So far, there is not a polynomial time algorithm run on classical computers which can be used to compute $\mbox{ord}_n(x)$. 
In 1994, P. Shor \cite{S97} proposed a quantum  algorithm which is claimed to be possible to compute $\mbox{ord}_n(x)$ in polynomial time.

The Shor's algorithm  uses two quantum
registers. It  needs an efficient quantum modular exponentiation method.
Recently  we \cite{CL14} have  found that the Shor's algorithm has to invoke the  unitary operation $U$ with $O(q^2)$ times, 
which can not be implemented in polynomial time, where $n^2 \leq  q < 2n^2$, $n$ is the large number to be factored, $U|y\rangle\equiv |xy (\mbox{mod}\, n)\rangle$, $y\in\{0, 1\}^{\ell}$,  $\ell$ is the bit length of $n$.  So far, there are few literatures to investigate the  mysterious process $$ \frac 1{\sqrt q}\sum_{a=0}^{q-1}|a\rangle|0\rangle \rightarrow \frac 1{\sqrt q}\sum_{a=0}^{q-1}
|a\rangle|x^a(\mbox{mod}\, n)\rangle. $$  
 We have reported the flaw to some researchers, but only received a comment made by MIT professor Scott Aaronson. He  explained that (personal communication, 2014/09/02): \\ \vspace*{-1mm}
  
 \hspace*{8mm}\fbox{\shortstack[l]{
 The repeated squaring algorithm works (and works in polynomial time)\\
for any single $|a\rangle|0\rangle$, mapping it to $|a\rangle|x^a \,(\mbox{mod}\, n)\rangle$. But, because of the\\
 linearity of quantum
mechanics, this immediately implies that the algorithm\\
 must also work
for any superposition of $|a\rangle$'s, mapping $\sum_a |a\rangle$ to $\sum_a |a\rangle |x^a \,(\mbox{mod}\, n)\rangle$. }} \vspace*{3mm}\\
We do not think that his answer is convincing, because it is too vague to specify  \emph{how many and what quantum gates or unitary operations are used on each qubit or a group of qubits in the second quantum register} (see Ref.\cite{CL14}).
Except the the above flaw,  Shor's algorithm relates to a peculiar entanglement which has not yet been mentioned and investigated. 

At the end of the  Shor's  factoring  algorithm,
one should observe the first register and denote the measured result as an integer $c$. Its complexity argument comprises:
\begin{itemize}
\item[(1)]
The probability $p$ of seeing a quantum state $ |c, x^k \,(\mbox{mod}\, n)\rangle$ such that $r/2
\geq \{rc\}_q$ is greater than $ 1/{3r^2}$, where $n$ is the integer to be factored, $q$ is a power of 2 satisfying $n^2 \leq  q < 2n^2$
and $r=\mbox{ord}_n(x)$. For convenience, the notation $\mbox{mod}\, n$ will be omitted henceforth.
\item[(2)] There are $\phi(r)$ possible $c$ which can be used to compute the order $r$.
\item[(3)] The measured number in the second register, i.e., $x^k$,  takes  $r$
possible values $1, x, x^2, \cdots, x^{r-1}$.
\item[(4)] The success
probability of running the algorithm once is greater than  $r\cdot \phi(r)\cdot  \frac 1{3r^2}$.
Since $\phi(r)/r> \xi/\log\log r$ for some constant $\xi$, it concludes that the algorithm runs in polynomial time.
\end{itemize}
 Notice that  the complexity argument views $p$ as the joint probability
$\mbox{Pr}\,(X=c,
Y=x^k ),$ instead of the conditional probability $\mbox{Pr}\,(X=c\,|\,
Y=x^k ),$   where two random variables $X$ and $Y$  assume respectively  values  from the sets $\{0, 1, \cdots, q-1\}$ and $
\{1, x, \cdots, x^{r-1}\}$.

In  the extended abstract of this paper, we \cite{CL09} introduced a more quantum register into Shor's algorithm.  On the one hand, by Shor's argument we  proved
that
$$\mbox{Pr}\,(X=c, Y=x^k, Z=x^k)=\mbox{Pr}\,(X=c, Y=x^k). $$
On the other hand,  if the observed values in the latter two quantum  registers are random,  we have
$$
\mbox{Pr}\,(X=c, Y=x^k, Z=x^k ) < \mbox{Pr}\,(X=c, Y=x^k, Z=x^l), \quad \mbox{where}\ k\neq l.\\
$$
 But this contradicts that \begin{eqnarray*}
  \mbox{Pr}\,(X=c, Y=x^k)
&\geq& \mbox{Pr}\,(X=c, Y=x^k)\cdot \mbox{Pr}(Z=x^l|\{X=c, Y=x^k\})\\
&=&  \mbox{Pr}\,(X=c, Y=x^k, Z=x^l)
\end{eqnarray*}  where $\mbox{Pr}\,(Z=x^l|\{X=c, Y=x^k\})
$ is the conditional probability.
 It shows  that  the   measured  numbers in two quantum registers of the same
pre-measurement state should be  equal   if
Shor's complexity argument is sound.  However, this
  contradicts  the argument that there are $r$ different
observed values  for the second  register. So far, we have  only received  one anonymous comment on the extended abstract \cite{CL09}. It argues  that the states in the  three quantum  registers are entangled.

In this paper, we shall introduce  more
quantum registers into Shor's algorithm. We show that the measured numbers in the added
 quantum registers should still be equal to the measured number in the second register if Shor's factoring  algorithm runs in polynomial time.
 That is to say, all these quantum qubits in these registers should be entangled by the above comment.
 But we find that the entanglement involving many quantum registers can not be interpreted by the mechanism of EPR pairs and the like.
 Moreover, this is a peculiar entanglement which has never been mentioned and investigated.
 Since the complexity of Shor's factoring algorithm  depends essentially on the peculiar entanglement,
we think the claim that Shor's  algorithm runs in polynomial time is not doubtless from a theoretical point of view.

\section{Preliminary}

A quantum analogue of a classical computer operates with quantum bits involving quantum states. The state of a quantum computer is described as a basis vector in a Hilbert space.
A qubit is a quantum state $|\Psi\rangle $ of the form
$$|\Psi\rangle= a|0\rangle+b|1\rangle, $$
where the amplitudes $a, b\in \mathbb{C}$ such that $|a|^2+|b|^2=1,$  $|0\rangle$ and $|1\rangle$ are basis vectors of the Hilbert space. Here, the \emph{ket} notation $|x\rangle$ means that $x$ is a quantum state.
 The state of a quantum system having $n$ qubits is a
point in a $2^n$-dimensional vector space. Given a state
$$\sum_{i=0}^{2^n-1} a_i|\chi_i\rangle, $$ where the amplitudes are complex numbers such that $\sum_{i=0}^{2^n-1}|a_i|^2=1$ and each
$|\chi_i\rangle$ is a basis vector of the Hilbert space, if the machine is measured with respect to this basis, the probability of seeing basis state $|\chi_i\rangle$ is $|a_i|^2$.

Two quantum mechanical systems are combined using the tensor product.
For example,   a system of two qubits $|\Psi\rangle= a_1|0\rangle+a_2|1\rangle $ and $|\Phi \rangle= b_1|0\rangle+b_2|1\rangle $ can be written as
$$ |\Psi\rangle|\Phi \rangle= {a_1 \choose a_2}\otimes {b_1 \choose b_2}=
\left(\begin{array}{c}
   a_1b_1 \\
  a_1b_2 \\
  a_2b_1 \\
  a_2b_2 \\
 \end{array}
\right)
  $$
  We shall also use the shorthand notations  $|\Psi, \Phi \rangle$. We call a quantum  state having two or more components  \emph{entangled} state, if
it is  not a product state. According to the Copenhagen interpretation of quantum mechanics, measurement causes an instantaneous collapse of the wave function describing the quantum system into an eigenstate of the observable state that was measured.
If entangled, one object cannot be fully described without considering the other(s).

The entangled state $$\frac{|01\rangle+|10\rangle}{\sqrt 2} $$
is commonly referred to as an EPR pair, after Einstein, Podolsky
and Rosen who tried to use it to prove Quantum Mechanics incomplete.
 Suppose that the first qubit of the state
above is given to Alice and  the second one is given to Bob.
 When Alice measures her qubit in the computational basis, she will obtain
the outcomes $|0\rangle$ or $|1\rangle$ with equal probability.
 Thus, the system shall collapse to either $|01\rangle$ or $|10\rangle$ because the two qubits are
entangled. This means that, if
Bob measures his qubit afterwards, his outcome is completely determined by
Alice's measurement result.

 In short, EPR pair is a form of quantum superposition.
 When a measurement is made and it causes one member of such a pair to take on a definite value,
 the other member of this entangled pair will at any subsequent time  be found to have taken the appropriately correlated value.
 Thus, there is a correlation between the results of measurements performed on entangled pairs, and this correlation is observed even though the entangled pair may have been separated by arbitrarily large distances.

\section{Description of Shor's factoring algorithm}

  Shor's factoring algorithm proceeds as follows \cite{S97}. At the beginning of the algorithm, one has to
  find $q=2^s$ for some integer $s$ such that $n^2 \leq  q < 2n^2$, where $n$ is to be factored.

\begin{itemize}
\item[] \emph{Initialization}. Put register-1 in the following uniform superposition
$$ \frac 1{\sqrt q}\sum_{a=0}^{q-1}|a\rangle|0\rangle. $$

\item[] \emph{Computation}. Keep $a$ in
  register-1 and
  compute $x^a$ in register-2 for some randomly chosen integer $x$.  We then have the following state
$$ \frac 1{\sqrt q}\sum_{a=0}^{q-1}
|a\rangle|x^a\rangle. $$

\item[] \emph{Fourier transformation}.   Performing Fourier transform on register-1 \cite{NC00,YS07}, we obtain the following state
$$ \frac 1{q}\sum_{a=0}^{q-1}\sum_{c=0}^{q-1}
\mbox{exp}(2\pi iac/q)|c\rangle|x^a\rangle.
$$

\item[] \emph{Observation}.  It suffices  to
observe the first register.  The probability $p$ that the machine reaches the  state $|c, x^k\rangle$ is
$$  \left| \frac{1}{q}\, \sum_{a:\, x^a \equiv x^k} \mbox{exp}(2\pi i ac/q) \right|^2 \eqno(1) $$
where $ 0\leq k< r=\mbox{ord}_n(x) $,  the sum is over all $a\, (0 \leq a < q)$  such that $x^a\equiv
x^k$. 

\item[] \emph{Continued fraction expansion}.
If there is a $d$ such that $$\frac{-r}2 \leq dq-rc\leq \frac r 2  $$
then the probability of seeing $|c, x^k\rangle$ is greater than $1/3r^2$ (see the following section for the argument of this claim).
Hence, we have
$$ \left|\frac d r -\frac c q\right|\leq \frac 1 {2q}$$
Since  $q\geq  n^2$,  we can round $c/q$ to obtain  $d/r$. Thus  $r$ can be obtained.
\end{itemize}

\section{The complexity argument of Shor's factoring algorithm}
Here is a brief description of the complexity argument of Shor's factoring algorithm. We refer to Ref.\cite{S97} for details.
 Setting $a=b r+k$ for some integer $b$ and the order $r=\mbox{ord}_n(x)$,  the probability $p$  is
$$\left|\frac 1{q}\sum_{b=0}^{\lfloor(q-k-1)/r \rfloor}e^{2\pi i(br+k)c /q}
 \right|^2.  $$
Then it argues that the probability $p$  equals to
$$\left|\frac 1{q}\sum_{b=0}^{\lfloor(q-k-1)/r \rfloor}e^{2\pi ib\{rc \}_q/q}
 \right|^2.  $$
Writing the above sum into an integral, we
obtain $$ \frac 1{q}
\int_{b=0}^{\lfloor\frac{(q-k-1)}{r} \rfloor}e^{2\pi
ib\{rc\}_q/q}\mbox{d} b \\
+ O\left( \frac{\lfloor (q-k-1)/r \rfloor}{q}(e^{2\pi i
\{rc\}_q/q}-1)\right).
 $$
  Taking $u =
rb/q$, we have
$$\frac {1}{r} \int_0^{\frac{r}{q} \lfloor \frac{q-k-1}{r}\rfloor} \mbox{exp}
\left(2\pi i u \frac{\{rc\}_q}{r} \right)\mbox{d}u $$ Taking into account that $k < r$,
  we can obtain the approximation $$\frac {1}{r} \int_0^{1} \mbox{exp}
\left(2\pi i u \frac{\{rc\}_q}{r}  \right)\mbox{d} u. $$
Hence, we have
$$ \left|\frac {1}{r} \int_0^{1} \mbox{exp}
\left(2\pi i u \frac{\{rc \}_q}{r} \right)\mbox{d} u \right|\geq 2/(\pi r)$$
 Therefore,
  $$p \geq \frac{4}{\pi^2 r^2}>1/3r^2.$$
Since $r=\mbox{ord}_n(x)$,  there are $r$ different  $x^{k}\mod  n$. If $(d, r)=1$, there are  $\phi(r)$ different $d$
 where $\phi$ is Euler's totient function
\cite{HW79,K81}.  Thus, the probability of obtaining $r$ is greater than $ \phi(r)\cdot r \cdot 1/3r^2=\phi(r)/3r$. This leads to a polynomial time algorithm for factorization.

For convenience, we now depict the Shor's complexity argument by Graph-1.

\hspace*{8mm}\begin{minipage}{\linewidth}
\includegraphics[angle=0,height=6.5cm,width=12cm]{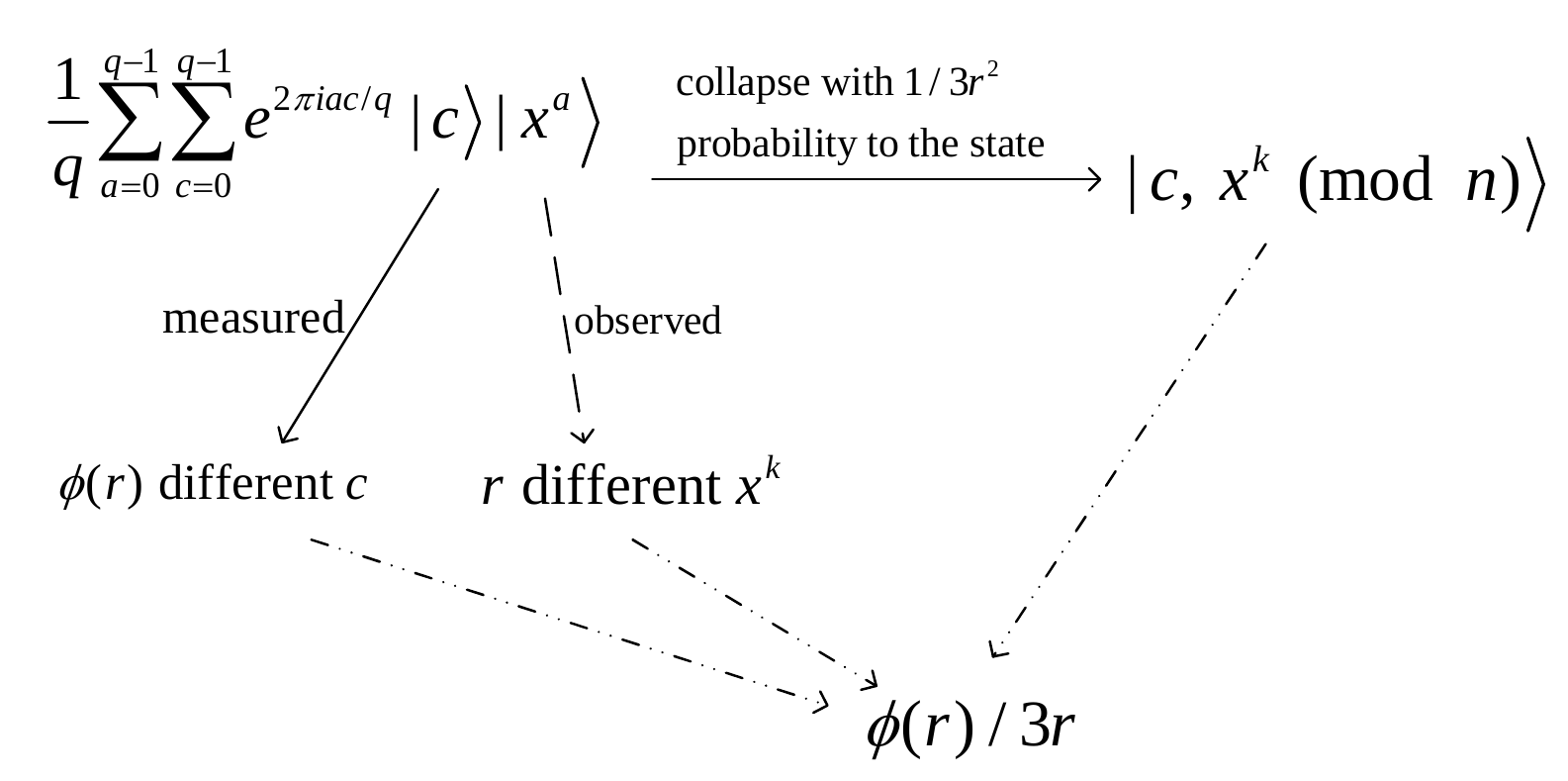}
 \end{minipage}

\centerline{\textsf{Graph-1: Shor's complexity argument}}

\section{On the lower bound to the
  joint probability}

It is easy to find that P. Shor  views
$1/3r^2$ as the lower bound to the joint  probability $P(X=c,
Y=x^k)$, where $r/2\geq \{rc\}_q$, the random variables $X$ and $Y$ values respectively
  from the sets  $\{0, 1, \cdots, q-1\} $ and  $\{1, x, \cdots,
x^{r-1} \}$.   Frankly speaking, it is difficult to  argue mathematically whether the expression of
Eq.(1) is a joint probability or not.
 \emph{But we here stress that  the expression of
Eq.(1) is directly defined over $x^k$ which is the observed value in the second register}. We shall argue that it is better to view  $p$  as
 the conditional probability $P(X=c \,| \, Y=x^k )$ rather than the joint probability
  $P(X=c, Y=x^k )$. The basic idea behind our argument is to investigate  a variation of  Shor's factoring algorithm which requires three quantum registers.

   Given an integer $n$ which is to be factored, find $q=2^s$ for some integer $s$ such that $n^2 \leq  q < 2n^2$.
\begin{itemize}
\item[] [Initialization] Put register-1  in the following uniform superposition state
$$ \frac 1{\sqrt q}\sum_{a=0}^{q-1}|a\rangle|0|0\rangle. $$

\item[] [Computation] Keep $a$ in
  register-1 and
  compute $x^{a}$ in register-2 and register-3 for some randomly chosen integer $x$, $1<x<n$.  The state becomes
$$ \frac 1{\sqrt q}\sum_{a=0}^{q-1}
|a\rangle|x^a|x^a\rangle. $$

\item[] [Fourier transformation] Perform Fourier transform on register-1. The state becomes
$$ \frac 1{q}\sum_{a=0}^{q-1}\sum_{c=0}^{q-1}
\mbox{exp}(2\pi iac/q)|c\rangle|x^a|x^a\rangle.
$$

\item[] [Observation]  Observe register-1.
The probability $p$ that the machine ends in the  state $|c, x^k, x^k\rangle$ is
$$  \left| \frac{1}{q}\, \sum_{a:\, x^a \equiv x^k} \mbox{exp}(2\pi i ac/q) \right|^2 \eqno(1') $$
where $ 0\leq k< r=\mbox{ord}_n(x) $,   the sum is over all $a\, (0 \leq a < q)$  such that $x^a\equiv
x^k$.

\end{itemize}

  To see the similarities of Shor's algorithm
with two registers and the variation with three registers, we refer to the
following Table-1.

\begin{center}
\begin{tabular}{|l|l|}
  \hline
 Shor's algorithm & A variation of Shor's algorithm  \\
  with two registers & with three registers \\
\hline (I)  \textit{Put} register-1 in the uniform  &
   (I) \textit{Put} register-1 in the uniform \\
  superposition. The state becomes &    superposition. The state becomes \\
\hspace*{4mm} $ \frac 1{\sqrt q}\sum_{a=0}^{q-1}|a\rangle|0\rangle
$ &\hspace*{4mm}  $ \frac 1{\sqrt q}\sum_{a=0}^{q-1}|a\rangle|0\rangle |0\rangle $ \\
(II) \textit{Compute} $x^{a} \, (\mbox{mod}\, n)$ in register-2.
 & (II)\textit{ Compute} $x^{a} \, (\mbox{mod}\, n)$ in register-2.  \\
 The state becomes   &
 and register-3. The state becomes\\
 \hspace*{4mm} $ \frac 1{\sqrt q}\sum_{a=0}^{q-1} |a\rangle|x^{a} \rangle $&
 \hspace*{4mm}  $ \frac 1{\sqrt q}\sum_{a=0}^{q-1}
|a\rangle|x^{a} \rangle|x^{a} \rangle $\\      (III)\textit{ Perform}
Fourier
transform  on  &  (III) \textit{Perform}  Fourier transform on  \\
register-1. The state becomes &  register-1. The
state becomes \\
 \hspace*{4mm} $ \frac 1{q}\sum_{a=0}^{q-1}\sum_{c=0}^{q-1}
e^{2\pi iac/q}|c\rangle|x^{a}\rangle $  &  \hspace*{4mm}  $
\frac 1{q}\sum_{a=0}^{q-1}\sum_{c=0}^{q-1} e^{2\pi
iac/q}|c\rangle|x^{a}\rangle|x^{a}\rangle
$ \\
(IV)\textit{Observe} the machine and \textit{compute} the  &(IV)
\textit{Observe} the machine and \textit{compute} the
\\probability of seeing the state $| c, x^{k}\rangle$.  It is & probability of seeing
the state $|c, x^{k},
x^{k}\rangle$. It is  \\
\hspace*{4mm} $ \left| \frac{1}{q}\, \sum_{a:\, x^{a} \equiv x^{k}}
e^{2\pi i ac/q} \right|^2   $ & \hspace*{4mm} $
\left| \frac{1}{q}\, \sum_{a:\, x^{a} \equiv x^{k}} e^{2\pi i
ac/q} \right|^2   $
\\  \hline
\end{tabular}\vspace*{3mm}

  \centerline{Table-1: Similarities of Shor's factoring algorithm and its variation with three registers}
\end{center}

Since the Eq.$(1)$ is the same as  Eq.(1'), by Shor's argument we easily prove that
$$ \mbox{Pr}\,(X=c,
Y=x^k)= \mbox{Pr}\,(X=c, Y=x^k, Z=x^k) \eqno(2) $$
However,  if the measured results in register-2 and register-3 are \textit{random}, then
$$
 \mbox{Pr}\,(X=c, Y=x^k, Z=x^k )<\mbox{Pr}\,(X=c, Y=x^k, Z=x^l), \eqno(3)$$
 where  $0\leq k, l \leq  r-1, k\neq l$.
 Thus,
\begin{eqnarray*}\mbox{Pr}\,(X=c, Y=x^k)&\geq &
\mbox{Pr}\,(X=c, Y=x^k)\cdot \mbox{Pr}\,(Z=x^l|\{X=c, Y=x^k\})
\\
&= &  \mbox{Pr}\,(X=c, Y=x^k, Z=x^l).
\end{eqnarray*}
  This is a contradiction.
  Notice that the contradiction originates directly from the Eq.(2).

So far, there are only two suggestions to resolve the above contradiction:
\begin{itemize}
\item[(I)] In our opinion, the expression
 $ \left| \frac{1}{q}\, \sum_{a:\, x^a \equiv x^k} \mbox{exp}(2\pi i ac/q) \right|^2  $ should be viewed as the conditional probability
 $\mbox{Pr}\,(X=c\,|\, Y=x^k) $, not the joint probability $\mbox{Pr}\,(X=c, Y=x^k)$.  In such case, we can not obtain Eq.(2).
 Moreover, the success probability of running Shor's factoring algorithm once is greater than ${\phi(r)}/{3 r^2}$, not ${\phi(r)}/{3 r}$.
 Thus, the claim that Shor's factoring algorithm  takes polynomial time is flawed.

 \item[(II)] The other suggestion claims  that the machine always ends in the state $|c, x^k, x^k\rangle $.
 It argues that all qubits in three quantum registers are entangled,  and  insists that the state $|c, x^k, x^l\rangle$ where $k\neq l$ can not be observed.
\end{itemize}

\section{Does the machine always end in the state $|c, x^k, x^k\rangle $}

There is a comment on the manuscript. It argues that:
 \emph{if the second register is exactly the same as the third register,
 then the probability of measuring two different numbers in the second
 and third registers is indeed zero.}
That is to say, the state $|c, x^k, x^l\rangle , k\neq l$ could not be observed  if
the latter two quantum registers have the same pre-measurement state. The
machine always ends in the  state $|c, x^k, x^k\rangle $. The argument is depicted by  Graph-2.
This argument insists that all qubits  in the three quantum registers are entangled.
 It claims that in  Graph-3 the probability $\alpha$ should be zero.
 
  By the argument,
if we introduce more quantum registers, we shall find that the measured numbers in the added
 quantum registers should still be equal to the the measured number in the second register¡¡(see  Table-2).
 In other words, the argument claims that all qubits in  those quantum registers should be entangled.

\hspace*{8mm}\begin{minipage}{\linewidth}
\includegraphics[angle=0,height=7cm,width=12cm]{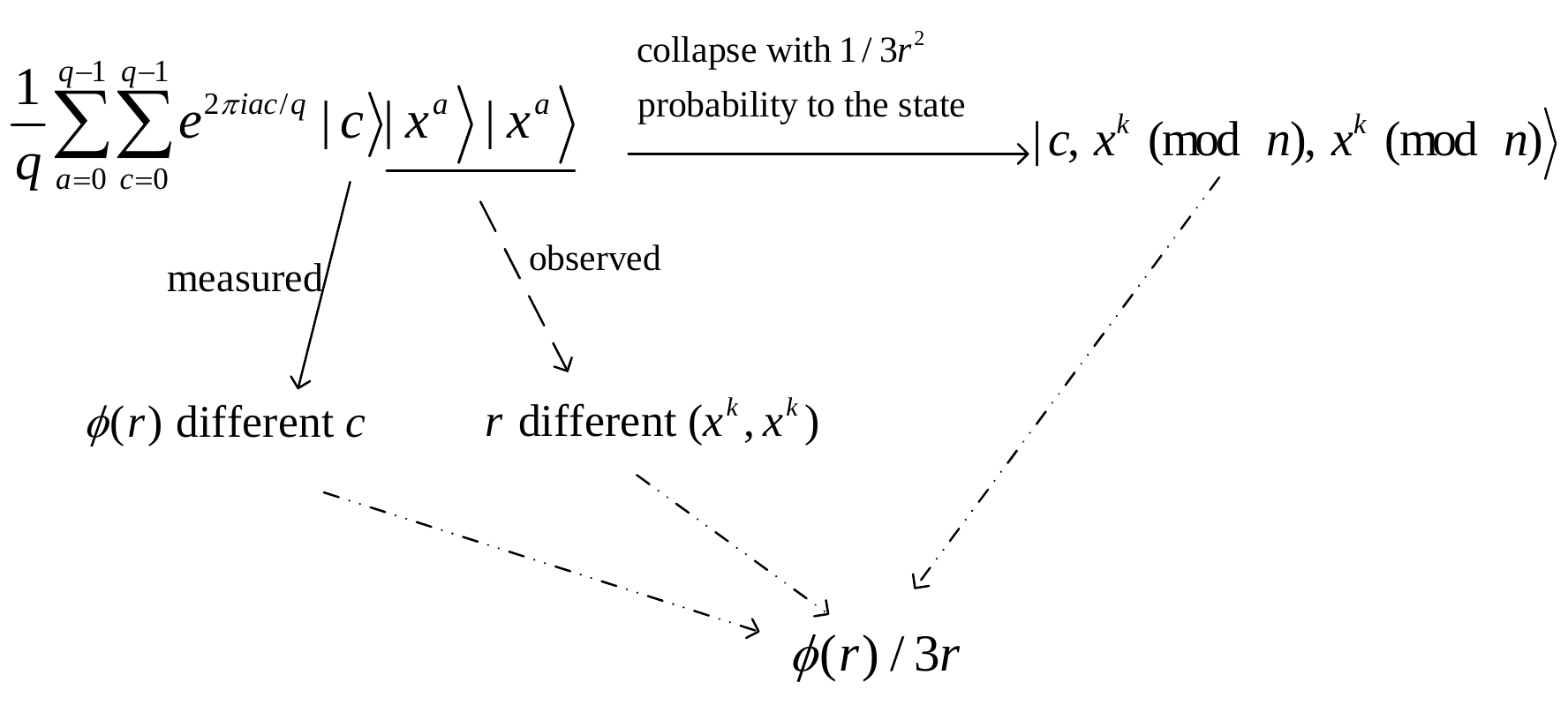}
 \end{minipage}

 \centerline{\textsf{Graph-2: The argument that the machine always ends in the state $|c, x^k, x^k\rangle $}}\vspace*{5mm}

 \hspace*{8mm}\begin{minipage}{\linewidth}
\includegraphics[angle=0,height=7cm,width=12cm]{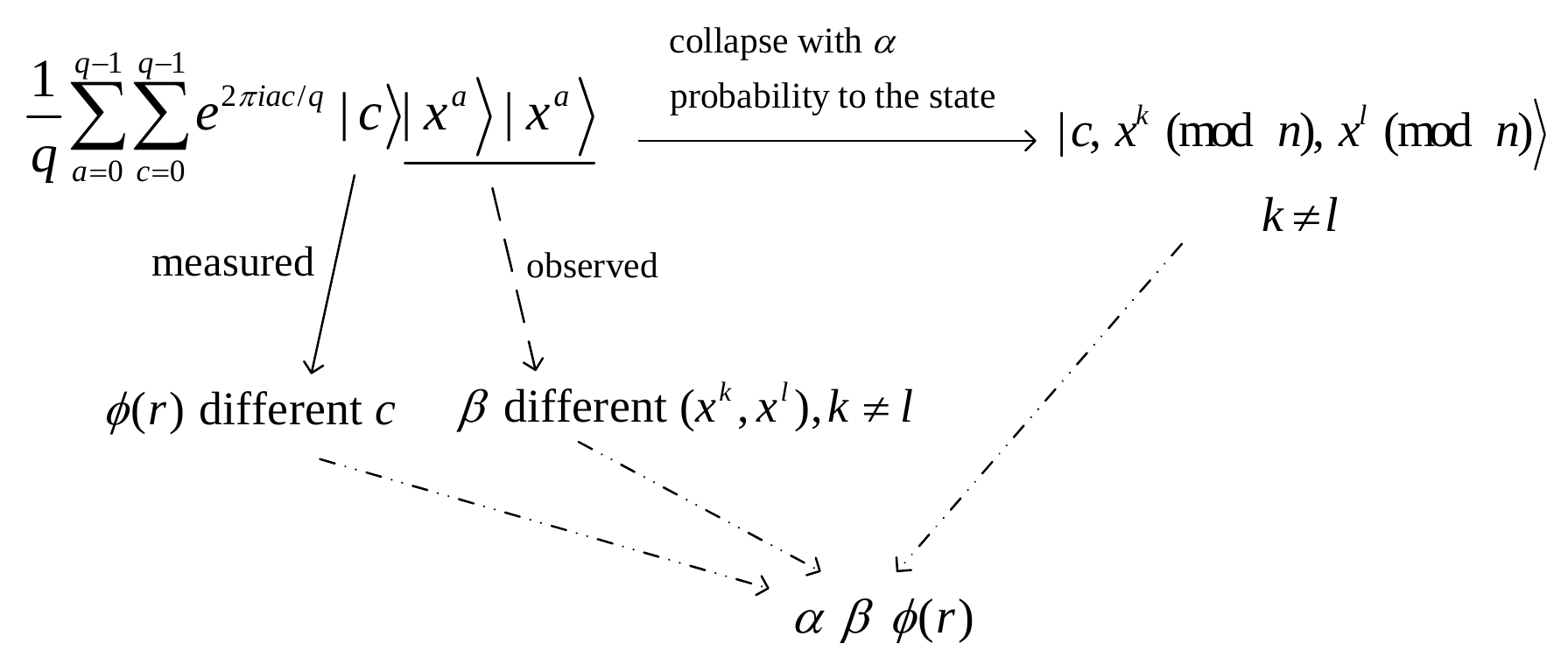}
 \end{minipage}

 \centerline{\textsf{Graph-3: The probability that the machine ends in the state $|c, x^k, x^l\rangle, k\neq l $}}\vspace*{5mm}

If the peculiar entanglement is indispensable for Shor's factoring algorithm, we should point out that:
\begin{itemize}
 \item[(1)] The entanglement involving many quantum registers has not yet been mentioned and investigated.
 All literatures related to Shor's factoring algorithm, such as Ref.\cite{SSB05,KM06,L07}, did not consider the peculiar entanglement.
  In 2006,  V. Kendon   and W.  Munro  \cite{KM06}
 investigated the topic that entanglement and its role in Shor's algorithm. They point out that:
 \begin{quotation}
 \noindent \emph{The nature of entanglement generated in the process of running Shor's factoring algorithm is still not fully understood, and little has been said about what role entanglement actually plays in quantum computation.}
 \end{quotation}  They \cite{KM06} used ``numerical simulation" to investigate how entanglement between register qubits
varies as Shor's algorithm is run on a quantum computer. They concluded that the inverse quantum Fourier transform can only generate entanglement within the upper register, or,
move entanglement around between the upper register qubits.

\item[(2)] It seems impossible to interpret the event (the measured numbers in two quantum registers with the same pre-measurement state should be equal) by the mechanism of EPR pairs and the like. Intuitively, the phenomenon violates Heisenberg's uncertainty principle.
     If the peculiar entanglement is indeed present during the course of Shor's algorithm, how to define the entanglement and how to certify it?

 \item[(3)] Surprisingly, the term ``entanglement" does not appear in Shor's paper \cite{S97}. It seems that the inventor did not realize at that time that
  his complexity argument of the famous algorithm  must rely on the peculiar entanglement.
 \end{itemize}

In summary, the claim that  the machine always ends in the state,
$$|c, \underbrace{x^k, \cdots, x^k}_{\ell}\rangle $$  in our opinion,  is not convincing because it
directly contradicts Shor's argument that there are $r$ different
observed values for register-2. 

 \begin{center}
{\small
\begin{tabular}{|l|l|}
  \hline
 Shor's algorithm& A variation of Shor's algorithm  \\
  with two registers  &   with $\ell+1$ registers \\
\hline (I)  \textit{Put}  register-1 in the uniform  &
   (I) \textit{Put}   register-1 in the uniform \\
superposition. The state becomes &    superposition. The state becomes \\
\hspace*{4mm} $ \frac 1{\sqrt q}\sum_{a=0}^{q-1}|a\rangle|0\rangle
$ &\hspace*{4mm}  $ \frac 1{\sqrt q}\sum_{a=0}^{q-1}|a\rangle\underbrace{|0\rangle\cdots  |0\rangle}_{\ell} $ \\
(II) \textit{Compute} $x^{a}$ in
register-2. & (II) \textit{Compute} $x^{a}$ in the latter $\ell$  \\
The state is   &
  registers. The state is\\
 \hspace*{4mm} $ \frac 1{\sqrt q}\sum_{a=0}^{q-1} |a\rangle|x^{a} \rangle $&
 \hspace*{4mm}  $ \frac 1{\sqrt q}\sum_{a=0}^{q-1}
|a\rangle\underbrace{|x^{a}\rangle \cdots|x^{a} \rangle}_{\ell} $\\  (III) \textit{Perform}
Fourier
transform  on  &  (III) \textit{Perform}   Fourier transform on  \\
  register-1. The state becomes &    register-1. The
state becomes \\
 \hspace*{4mm} $ \frac 1{q}\sum_{a=0}^{q-1}\sum_{c=0}^{q-1}
e^{2\pi i ac/q}|c\rangle|x^{a}\rangle $  &  \hspace*{4mm}  $
\frac 1{q}\sum_{a=0}^{q-1}\sum_{c=0}^{q-1} e^{2\pi
iac /q}|c\rangle\underbrace{|x^{a}\rangle\cdots|x^{a}\rangle}_{\ell}
$ \\
(IV) \textit{Observe} the machine and \textit{compute} the  &(IV)
\textit{Observe} the machine and \textit{compute} the
\\probability of seeing the state $|x, x^k\rangle$.  It is & probability of seeing
the state $|c, \underbrace{x^{k}, \cdots
x^{k}}_{\ell}\rangle$. It is  \\
¡¡\hspace*{4mm} $ \left| \frac{1}{q}\, \sum_{a:\, x^{a} \equiv x^{k}}
e^{2\pi i ac/q} \right|^2   $ & \hspace*{4mm} $
\left| \frac{1}{q}\, \sum_{a:\, x^{a} \equiv x^{k}} e^{2\pi i
ac /q} \right|^2  $
\\  \hline
\end{tabular}\vspace*{3mm}

 \centerline{Table-2:  A variation of Shor's algorithm with $\ell+1$ registers}
 }
\end{center}

\section{The latest experiments on  entanglement}

In 2012, it \cite{M12} reported a new experiment on generating,   manipulating and measuring  entanglement.
In the experiment
  \emph{four photons} are used.
  ``It can delay the choice of measurement-implemented through a high-speed tunable bipartite-state analyser and a quantum random-number generator-on two of the photons into the time-like future of the registration of the other two photons.
  This effectively projects the two already registered photons onto one of two mutually exclusive quantum states in which the photons are either entangled (quantum correlations) or separable (classical correlations)."

In another experiment, P. Shadbolt, et al. \cite{S12} reported
an integrated waveguide device that can generate and completely characterize pure \emph{two-photon}
states with any amount of entanglement and arbitrary single-photon states with any amount of
mixture.

 \section{Using Shor's factoring algorithm to certify a quantum computer}

 In 2001, Shor's algorithm was demonstrated by a group at IBM, who factored 15 into $3 \times 5$.  However, some doubts that whether IBM's experiment was a true demonstration of quantum computation have been raised, since no entanglement was observed.

 In 2012, E. Lucero et al. \cite{L12} claimed that they have run a three-qubit compiled version of
Shor's algorithm to factor the number 15. They  produce  coherent
interactions between five qubits and verify bi- and tripartite entanglement via quantum state tomography.

In the past ten years, it is somewhat depressing that several demonstrations of Shor's factoring algorithm only claimed to be able to factor 15 or 21.  It is even worse that almost  these demonstrations are criticised for lack of more scrutiny.

 On May 11, 2011, D-Wave Systems \cite{D} announced the D-Wave One,
   labeled ``the world's first commercially available quantum computer." The company claims this system uses a 128 qubit processor chipset.
   In early 2012 D-Wave Systems revealed a 512-qubit quantum computer.  In January 2014, researchers at UC Berkeley and IBM published a classical model explaining the D-Wave machine's observed behavior, suggesting that it may not be a quantum computer \cite{W14}. In May 2014, researchers at D-Wave, Google, USC, Simon Fraser University, and National Research Tomsk Polytechnic University published a paper containing experimental results that demonstrated the presence of entanglement among D-Wave qubits \cite{L14}.

 Historically speaking,  Shor's factoring algorithm is the main impetus of  developing quantum computers. In some senses,
 a computer could be called a ``quantum computer" only if it is able to factorize large integers by running Shor's factoring algorithm. We believe it is a universally acceptable standard to certify a quantum computer using  Shor's factoring algorithm. But it is a pity,  D-Wave has no intention to
 use its quantum computers to run the famous algorithm so as to certify itself.

\section{Conclusion}

 In this paper, by investigating Shor's factoring algorithm
  with more registers, we show that the measured numbers in many quantum registers with the same
pre-measurement state should be equal if the original argument is sound.
 We think the claim that Shor's factoring algorithm runs in polynomial time needs more investigations, especially,
 physical verifications for the peculiar phenomenon.
 We also throw some light on the question of what a quantum computer was like.

\textbf{Acknowledgements}. This work was supported by the National Natural Science Foundation of China (Grant Nos. 60970110, 60972034),
and the State Key Program of National Natural Science of China (Grant No. 61033014).


\begin{thebibliography}{11}

\renewcommand{\baselinestretch}{.9}
  \renewcommand{\arraystretch}{.9}
  \normalsize \small \parskip 0mm

\bibitem{M76} Miller G.:
Riemann's hypothesis and tests for primality. J. Comput. System
Sci., 13: 300-317 (1976)

\bibitem{S97} Shor P.: Polynomial-time algorithms for prime factorization and discrete
logarithms on a quantum computer. SIAM J. Comput.  26 (5): 1484-1509 (1997)

\bibitem{CL14} Cao ZJ. Liu LH: A Note on the Quantum Modular Exponentiation Method
 Used in Shor's Factoring Algorithm. \verb"http://arxiv.org/abs/1408.6252v1" (2014)
 
\bibitem{CL09} Cao ZJ. Liu LH: On the Complexity of Shor's Algorithm for Factorization. In:
 Proceeding of 2nd International Symposium on Information Science and Engineering, pp.164-168. IEEE (2009)
 
\bibitem{NC00} Nielsen M., and
  Chuang I.:  Quantum Computation and Quantum Information. Cambridge
University Press (2000)

\bibitem{YS07} Yoran N., Short A.:  Efficient classical simulation of the
approximate quantum Fourier  transform. Phys.Rev.A.76.042321 (2007)

\bibitem{HW79} Hardy G., Wright E.:
 An Introduction to the Theory of Numbers, Fifth ed., Oxford
University Press, New York (1979)

\bibitem{K81} Knuth D.: The Art of Computer Programming,
Vol. 2: Seminumerical Algorithms, Second ed., Addison-Wesley (1981)

\bibitem{SSB05} Shimoni Y.,  Shapira D.,  Biham O.:  Entangled quantum states
generated by Shor's factoring algorithm. Phys. Rev. A 72, 062308
(2005)

\bibitem{KM06}  Kendon  V., Munro W.: Entanglement and its Role in Shor's
 Algorithm. Quantum Information and Computation, 6 (7), pp. 630-640 (2006)

\bibitem{L07}  Lanyon B.,  et al.:  Experimental Demonstration of a Compiled Version
of Shor's Algorithm with Quantum Entanglement.  Phys. Rev. Lett 99
(2007)

\bibitem{M12}  Ma X.S.,   Zotter S.,   Kofler J.,  Ursin R.,  Jennewein T.,  Brukner C.,  Zeilinger A.:  Experimental delayed-choice entanglement swapping. Nature Physics. \verb"doi:10.1038/nphys2294" (2012)

\bibitem{S12} Shadbolt P., et al.: Generating, manipulating and measuring entanglement and mixture with a reconfigurable photonic circuit. Nature Photonics 6: pp. 45-59 (2012)

\bibitem{L12}  Lucero E.: Computing prime factors with a Josephson phase qubit quantum processor. Nature Physics 8, pp.719-723 (2012)


\bibitem{D} \verb"http://en.wikipedia.org/wiki/D-Wave_Systems"

\bibitem{W14} Vinci W. et al.:  Distinguishing Classical and Quantum Models for the D-Wave Device,  \verb"http://arxiv.org/abs/1403.4228"

\bibitem{L14} Lanting T. et al.:  Entanglement in a Quantum Annealing Processor.  Physical Review X 4, 021041 (2014). \verb"https://journals.aps.org/prx/pdf/10.1103/PhysRevX.4.021041"
    
\end{thebibliography}
\end{document}